\documentclass[twocolumn,aps,prl,superscriptaddress,showpacs]{revtex4}
\usepackage{amsmath}
\usepackage{hyperref}
\usepackage{epsfig}
\usepackage{epsf}
\usepackage{array}
\usepackage{color}
\usepackage{graphicx}
\usepackage{epstopdf}
\usepackage{psfrag}
\usepackage{float}
\begin{document}
\bibliographystyle{apsrev}
\title{Two-particle response in Cluster Dynamical Mean-Field Theory: Formalism and application to the Raman Response of High-temperature Superconductors}
\author{Nan Lin}
\affiliation{Department of Physics, Columbia University, New York, New York 10027, USA}
\author{Emanuel Gull}
\affiliation{Max Planck Institute for the Physics of Complex Systems, Dresden, Germany}
\author{Andrew J. Millis}
\affiliation{Department of Physics, Columbia University, New York, New York 10027, USA}
\date{\today }

\begin{abstract}
A method is presented for the unbiased numerical computation of two-particle response functions of correlated electron materials via a solution of the dynamical mean-field equations in the presence of a perturbing field. The power of the method is demonstrated via a computation of the Raman $B_{1g}$ and $B_{2g}$ scattering intensities of the two dimensional Hubbard model, in parameter regimes believed to be relevant to high-temperature superconductivity. The theory reproduces the `two-magnon' peak characteristic of the Raman intensity of the insulating parent compounds of high-$T_c$ copper oxide superconductors and shows how it evolves to a quasiparticle response as carriers are added. The method can be applied in any situation where a solution of the equilibrium dynamical mean-field equations is feasible.
\end{abstract}
\maketitle
The development of dynamical mean-field theory, first in its single-site version \cite{Georges96} and then in cluster extensions \cite{Maier05}, along with its interface to band theory \cite{Kotliar06,Held06}, has transformed our understanding of correlated electron physics. In the particular case of the two-dimensional Hubbard model, believed \cite{Anderson87} to represent the physics of high temperature superconductors, the method has provided new insights into the correlation-driven (`Mott') metal-insulator transition \cite{Georges92,Gull10}, the pseudogap regime that separates the Mott insulator from the Fermi liquid metal \cite{Parcollet04,Park08,Gull08,Gull10} and the existence \cite{Lichtenstein00,Maier00,Maier05_dwave} and properties \cite{Haule07,Civelli08,Kancharla08,Gull12} of a d-wave superconducting state. However, dynamical mean-field theory is based on an approximation to the one-electron Green function $G(k,\omega)$, measurable in angle-resolved photoemission experiments \cite{Damascelli03} while theoretical analysis of wide classes of experiments including optical conductivity, Raman spectroscopy and inelastic neutron scattering requires  a vertex function whose computation has proven very challenging.
While the charge vertex can be obtained analytically in simplified situations such as the Falicov-Kimball model \cite{Shvaika05}, computations of the vertex for interacting electron models  have not, in practice, been carried out in full generality.  Kune\v{s}  \cite{Kunes11} obtained the zero frequency charge and spin  vertices corresponding to single-site dynamical mean-field theory of the Hubbard model, and Yang {\it et al.} \cite{Yang11} obtained the spin, charge and superconducting vertices for larger clusters. However,  calculations  \cite{Toschi07,Park11} of the full dynamic ($\omega\neq0$) response have been based on the single site approximation and have employed truncations of the general frequency dependence. 

In this paper we present a new method for determining the two-particle response in cluster dynamical mean-field theory and demonstrate its effectiveness via a computation of the doping dependence of the Raman scattering amplitude of the two-dimensional Hubbard model from the Mott insulating to the Fermi liquid regime. Raman spectroscopy has been of fundamental importance to high temperature superconductivity \cite{Devereaux07} but the theoretical description  in terms of an underlying Hubbard model involves vertex corrections in an essential way and has not been systematically studied. 

To introduce our method we recall salient features of the theory of linear response  \cite{Kadanoff62}, defined quantum mechanically as the leading-order difference of the  expectation value of an operator ${\mathbf{\hat R}}$ in the presence and absence  of a probe field $P$. This is given by $\langle{\mathbf{\hat R}}\rangle_P =\text{Tr}\left[\mathbf{\hat R}\mathbf{G}_P\right]-\text{Tr}\left[\mathbf{\hat R}\mathbf{G}^{eq}\right]=
\chi_{RP}P+\mathcal{O}(P^2)$ with
\begin{equation}
\label{chidef}
\chi_{RP}=-\text{Tr}\left[\mathbf{\hat R}\mathbf{G}^\text{eq}\frac{\delta \mathbf{G}_0^{-1}}{\delta P}\mathbf G^\text{eq}\right]+\text{Tr}\left[\mathbf{\hat R}\mathbf G^\text{eq}\frac{\delta \mathbf{\Sigma}}{\delta P}\mathbf G^\text{eq}\right].
\end{equation}
Here $\mathbf G^\text{eq}=(\mathbf{G}_0^{-1}-\mathbf{\Sigma}^\text{eq})^{-1}$ is the equilibrium ($P=0$) Green function, related in the standard way to a bare Green function $\mathbf{G}_0$ and a self energy $\mathbf \Sigma$.  We have omitted a possible term arising from explicit dependence of $\mathbf{\hat R}$ on $P$; this gives rise to the `diamagnetic' term in the optical conductivity but is not otherwise relevant. The first term in Eq.~\ref{chidef} gives the `bubble term', computable from knowledge of the one-electron Green function and the bare vertex $\delta\mathbf{G}_0^{-1}/\delta P$; the second term, arising from changes in the many-body physics due to the perturbation, is the vertex correction of interest here.

For wide classes of strongly correlated materials, neither perturbative diagrammatic expansions about a mean-field solution  nor  partial (e.g RPA or GW)  resummations suffice; a fully nonperturbative  treatment is required. For the one electron Green function, cluster dynamical mean-field theory \cite{Maier05} provides such a treatment. In this theory the electron self energy $\mathbf \Sigma$,  a matrix  in the full single-particle Hilbert space of the problem, is approximated in terms of a much smaller number $N_c$ of functions $\Sigma^{\alpha\beta}$ as
\begin{equation}
\mathbf{\Sigma} \approx \mathbf{\Sigma}_\text{approx}=\sum_{\alpha,\beta=1...N_c}\mathbf{\phi}_{\alpha\beta}\Sigma^{\alpha\beta}.
\label{dmftformal}
\end{equation}
Different choices for $\mathbf{\phi}_{\alpha\beta}$ give different versions of dynamical mean-field theory while the $\Sigma^{\alpha\beta}$ are the self energies of a quantum impurity model specified by the interactions of the original model and by mean-field functions $\left(\mathcal{G}_0^{-1}\right)^{\alpha\beta}$ determined by the self consistency condition 
\begin{equation}
\left(\mathcal{G}_0^{-1}\right)^{\alpha\beta}=\Sigma^{\alpha\beta}+
\left(
\text{Tr}\left[
\mathbf{\phi}_{\alpha\beta} \left(
\mathbf{G}_0^{-1}-\mathbf{\Sigma}_\text{approx}
\right)^{-1}
\right]
\right)^{-1}\!\!\!\!\!.
\label{scegeneral}
\end{equation}
Here the trace  and inner inversion are over the single particle Hilbert space of the full problem and the outer inversion  is in the impurity model space. All quantities may depend on the perturbation $P$ and we compute $\delta \mathbf{\Sigma}/\delta P$ directly by solving  to first order in $P$.  

Linearization of Eq.~\ref{scegeneral} yields the first order correction  $(\mathcal{G}^{-1}_0)^{1}_{\alpha\beta}$  in terms of $\delta {\mathbf G}_0^{-1}/\delta P$ (known {\it a-priori}) and $\delta \Sigma^{\alpha\beta}/\delta P$ (to be computed). From  this and the impurity model four-point function $\Gamma$, written here for a monochromatic perturbation of frequency $\Omega$ as  
\begin{align}
\Gamma^{\Omega;\alpha\beta}_{\beta'\alpha'}(\omega,\omega') = \left\langle\! c_{\alpha}^{}(\omega+\Omega) c_{\beta}^\dagger(\omega) c_{\beta^{'}}^{}(\omega')c^{\dagger}_{\alpha{'}}(\omega'+\Omega)\!\right\rangle
\label{4point}
\end{align}
we obtain the first order correction  $\mathbf{G}^1$ to the impurity model Green function as
\begin{align}
G^1_{\alpha\beta}(\omega+\Omega,\omega) =
&T^2\!\!\!\!\sum_{\omega_1,\alpha{'}\beta{'}}\!\!\!\Gamma^{\Omega;\alpha\beta}_{\beta'\alpha'}(\omega,\omega_1)(\mathcal{G}_0^{-1})^1_{\alpha'\beta'} (\omega_1+\Omega,\omega_1)
 \nonumber\\
-\delta_{\Omega,0}G^{\rm eq}_{\alpha\beta}(\omega)&\!\!\!\!\sum_{\omega_1,\alpha'\beta'}\!\!\!G^{\rm eq}_{\alpha'\beta'}(\omega_1)(\mathcal{G}_0^{-1})^1_{\alpha'\beta'} (\omega_1,\omega_1).
\label{g1a1}
\end{align}
From this and the linearized impurity-model Dyson equation follows $\delta\Sigma^{\alpha\beta}/\delta P$. The resulting linear equation is solved for $\delta\Sigma^{\alpha\beta}/\delta P$, from which we  obtain the desired vertex correction $\delta \mathbf{\Sigma}/\delta P$ via Eq.~\ref{dmftformal}. 

Previous dynamical mean-field literature introduced \cite{Georges96,Jarrell01}, and used \cite{Toschi07,Kunes11,Park11,Rohringer12}, a  different approach, inverting the impurity model Bethe-Saltpeter equation to obtain the two-particle irreducible impurity-model vertex in terms of $\Gamma$ and then using this in the lattice Bethe-Saltpeter equation to compute the physical vertex. Our procedure replaces  the numerical inversion of the  Bethe-Saltpeter equation 
(which  requires considerable care \cite{Kunes11} to avoid numerical instabilities \cite{Slezak09,Toschi07,Rohringer12})
by the solution of the linearized DMFT equation (which we have found not to be problematic), and avoids the second Bethe-Saltpeter equation by constructing the fully reducible lattice vertex directly.

The key issue in the implementation is the measurement and storage of  $\Gamma$, which is needed in each sector for a wide range of $\omega,\omega_1$ at every relevant $\Omega$.  It is necessary to compute $\Gamma$ in a strip $|\omega-\omega_1|<\Delta_1$ and $|\omega+\omega_1+\Omega|<|\Omega|+\Delta_2$. $\Delta_{1,2}$ are increased empirically until the final $\chi$ ceases to change on the $\sim 1\%$ level and we typically need   $\Delta_1\sim \Delta_2/2$  of order 3-4 times a relevant frequency scale (interaction strength or bandwidth). $\Gamma$ is  measured on the imaginary time axis and the Fourier transform to frequencies is a significant portion of the computational burden.  Computations are substantially accelerated by a formal rearrangement which allows the needed information to be obtained from a small set of one dimensional Fourier transforms rather than a two dimensional one, and by recent extensions of the fast Fourier transform method to non-uniform grids \cite{Potts09}.  

\begin{figure}[tbh]
\centering
\includegraphics[width=0.95\columnwidth]{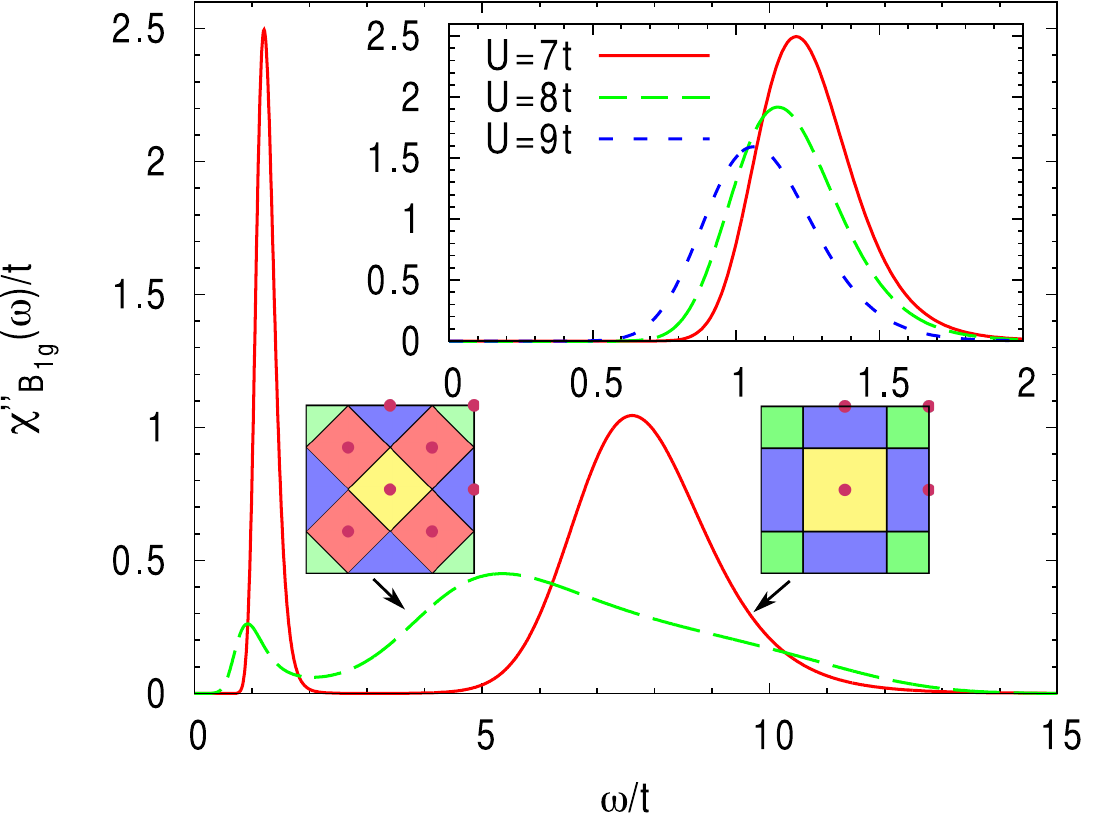}
\caption{Raman $B_{1g}$ scattering intensity in the 4-site (solid line, red online)  and 8-site (dashed line, green on-line) DCA approximations for parameters $U=7t$, $t'=-0.15t$, $\beta =10/t$ and $n=1$. Upper right inset: interaction strength dependence of $B_{1g}$ Raman intensity at density $n=1$ calculated for the four site cluster. Lower two insets: momentum space partitioning for the four- and eight-site cluster geometries considered in this paper. 
}
\label{cluster}
\end{figure}

To demonstrate the power of the new approach we compute the non-resonant Raman scattering intensity of  the two dimensional Hubbard model defined by
\begin{equation}
H_{\rm Hub} = H_0+U\sum_i n_{i\uparrow}n_{i\downarrow}.
\label{hubbard}
\end{equation}
For definiteness we take $H_0$ $=$ $\sum_{k\sigma}$ $(\varepsilon_k$  $-$ $\mu)$ $c^\dagger_{k\sigma}c^{}_{k\sigma}$ with $\varepsilon_k$ $=$ $-$ $2t$ $(\cos k_x$ $+$ $\cos k_y)$ $-$ $4t'$ $\cos k_x$ $\cos k_y$ with $t'$ $=$ $-0.15t$ and $U=7t$. To fix the energy scale we use $t=0.35\ {\rm eV}$, a value  generally accepted for high-$T_c$ superconductors \cite{Andersen95}.

We  consider two scattering geometries: $B_{1g}$, where the electric fields of the incident and outgoing photons are directed along the Brillouin zone axes ($k_x$ or $k_y=0$) highlighting the antinodal region, and $B_{2g},$ where the electric field vectors are directed along Brillouin zone diagonals ($k_x=\pm k_y$) highlighting the nodal region. The perturbing terms corresponding to the $B_{1g}$ and $B_{2g}$ scattering channels are (in the non-resonant approximation)

\begin{eqnarray}
H^{B_{1g}}_{\rm Raman}&=& P \left[\frac{1}{2} \left(\frac{\partial^2 H_0}{\partial k_x^2} - \frac{\partial^2 H_0}{\partial k_y^2}\right)\right], \\
H^{B_{2g}}_{\rm Raman}&=& P \frac{\partial H_0^2}{\partial k_x\partial k_y}.
\end{eqnarray}

We use  the dynamical cluster approximation (DCA) \cite{Maier05} in which the Brillouin zone of  momentum space is partitioned into $N_c$ equal area tiles labeled by central momentum $K$ and study in particular  $N_c=4$ and $8$ (see insets to Fig.~\ref{cluster}). Both lattice and impurity $\Sigma$ in Eq.~\ref{dmftformal} are diagonal. The lattice quantities depend on a continuous momentum $k$, the index $\alpha=\beta$ represents cluster momentum $K$ and $\phi^k_K=1$ if $k$ is in tile $K$ and zero otherwise so the lattice trace is just a momentum integral over the tile. We solve the model on the imaginary axis in the paramagnetic phase using the numerically exact continuous-time auxiliary field (CT-AUX) impurity solver \cite{Gull08b} with submatrix updates \cite{Gull11} and analytically  continue the final $\chi_{RP}(i\Omega_m)$ using the maximum entropy technique \cite{Jarrell96} taking into account covariance matrices estimated by a jackknife procedure. 

\begin{figure}[tbh]
\centering
\includegraphics[width=0.95\columnwidth]{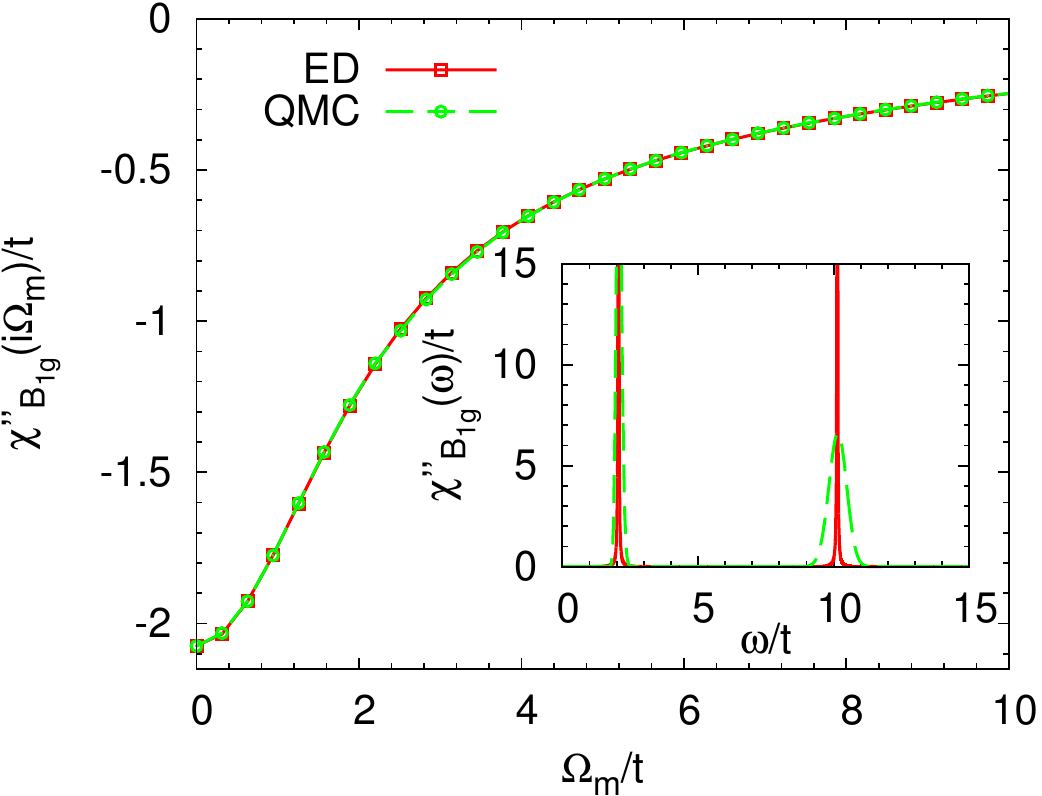}
\caption{Raman $B_{1g}$ scattering function  of isolated 4-site cluster calculated at $U=8t$ and $\beta=20/t$ by exact diagonalization (solid lines)  and Quantum Monte Carlo (dashed lines) for Matsubara (main panel) and real frequencies (inset). }
\label{isolated_U8}
\end{figure}

To test the formalism we observe that if the momentum integral is replaced by an evaluation at the central momentum $K$ then our procedure is reduced to the solution of an isolated cluster, which can also be solved by exact diagonalization (ED). The main panel of Fig.~\ref{isolated_U8} compares the  Matsubara axis Raman $B_{1g}$ scattering intensity obtained by applying our procedure  to the isolated four site cluster with that obtained from a direct diagonalization of the same isolated cluster. The results are seen to be identical up to Monte Carlo errors which are smaller than the symbol size. The inset compares the real axis spectrum obtained by analytical calculation to that obtained directly from the exact solution. Apart from a broadening, the two procedures give the same result; in particular the areas of the peaks are the same in the two methods.

Fig.~\ref{cluster} shows the calculated Raman $B_{1g}$ intensity  at carrier density $n=1$, i.e. in the paramagnetic Mott insulating phase. The spectra exhibit a two peak structure. The peak at  higher energies corresponds to  quasiparticle excitations across the Mott gap $\sim \omega=3t$ for the 8-site (integrated area $2.5$) and $\sim \omega=5t$ for the 4-site approximation (integrated area $3.1$). We identify the lower energy peak (area $1.0$ for 4-site and $0.2$ for 8-site)   as arising from the creation of a pair of spin flip excitations  because at $n=1$ and large $U$ these are the only excitations available in this energy range.   This identification is corroborated by computations (inset) of the interaction strength dependence in the 4-site cluster (present-day computational limits preclude study of $U>7t$ for the 8-site cluster): as $U$ is increased the upper feature shifts up in energy and the lower feature shifts down, as expected for a peak scaling with $J\sim t^2/U$. 

While the 4 and 8 site calculations are qualitatively similar, they are quantitatively different. In the DCA approximation the  4-site cluster is known \cite{Gull10} to have properties different from all of the other clusters, among other things strongly overestimating singlet formation and insulating behavior. In the rest of this paper we focus on the 8-site cluster, believed \cite{Gull10} to be more representative of the physics of the model. 

\begin{figure}[tbh]
\centering
\includegraphics[width=0.95\columnwidth]{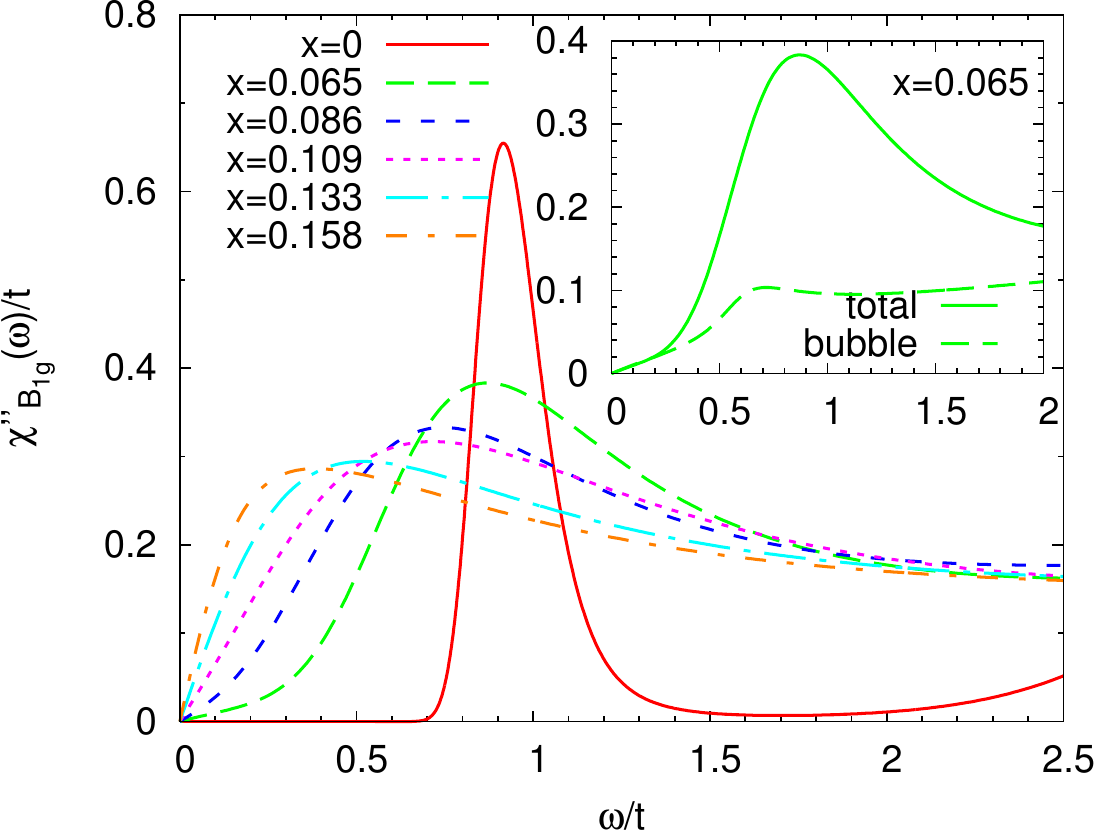}
\caption[Raman $B_{1g}$ scattering including vertex correction in 8-site DCA]{Main panel: Raman $B_{1g}$ scattering intensity in 8-site cluster for parameters $U=7t$, $t'=-0.15t$ and $\beta =20/t$ at  dopings indicated. 
Inset: separation into bubble and total contribution for $x=0.065$.
}
\label{8site_B1g_beta20}
\end{figure}

Fig.~\ref{8site_B1g_beta20} presents the evolution of the $B_{1g}$ spectra.  When the insulator is doped away from half-filling, the peak at $\omega \sim t$ broadens and shifts to lower frequency, while at lowest frequencies a component $\chi(\omega)\sim \omega$ appears and increases rapidly with doping. The inset decomposes the $x=0.065$ spectrum into bubble and vertex correction (visible as the difference between the two curves). The bubble diagram accounts for the entire quasiparticle part while the vertex correction, which  in the insulating state gives rise to the  two-spin-flip peak, produces the higher frequency maximum.  We therefore attribute the higher frequency peak to the relic of the two spin-flip peak in the metallic state and  the low frequency $\chi(\omega)\sim \omega$ feature to quasiparticles.  At the lowest dopings $x = 0.065$ and $x = 0.086$ the pseudogap is visible as a change in slope from the very low frequency regime (dominated by quasi- particles) to an intermediate energy regime where much of the scattering comes from the two spin-flip feature. As the doping is increased beyond $x=0.1$ the pseudogap disappears, the two-spin flip and quasiparticle scatterings merge, and the vertex correction decreases in importance, becoming completely negligible by doping $\sim 0.24$. These data are in excellent semiquantitative agreement with the  $B_{1g}$ spectra reported in Fig.~9 of Ref.~\onlinecite{Sugai03}.

\begin{figure}[tbh]
\centering
\includegraphics[width=0.95\columnwidth]{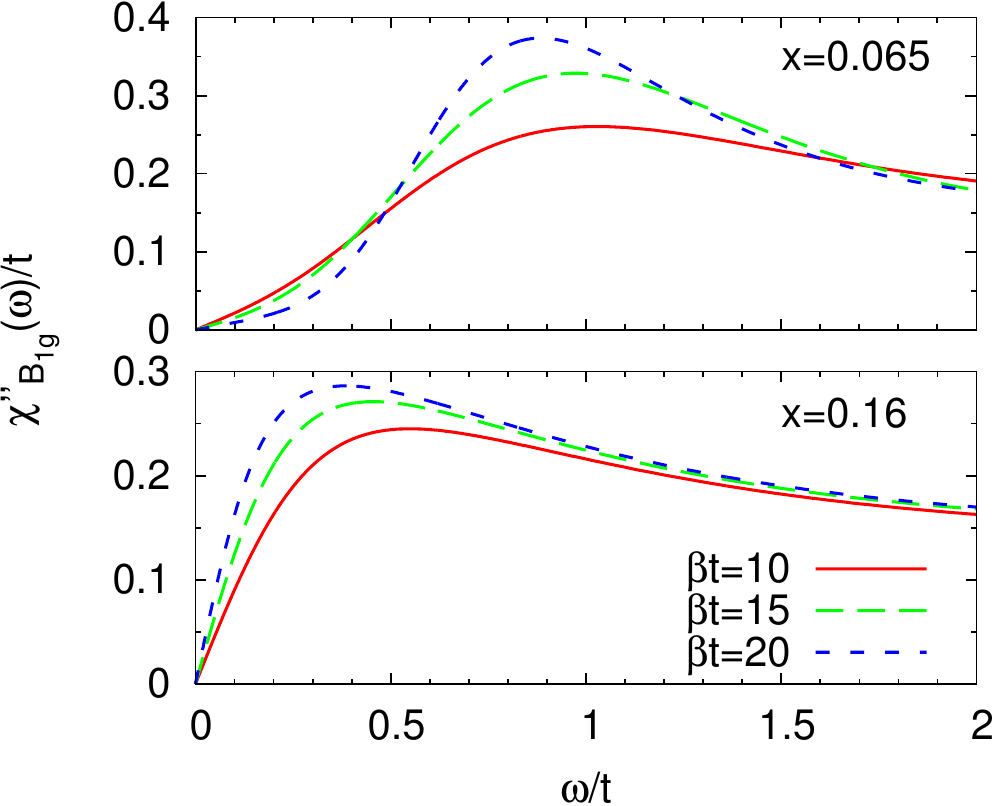}
\caption[Temperature dependence of the Raman $B_{1g}$ scattering in 8-site DCA]{Temperature dependence of Raman $B_{1g}$ scattering intensity in 8-site DCA at hole dopings indicated. }
\label{8site_B1g_doped_temp}
\end{figure}

Fig.~\ref{8site_B1g_doped_temp} shows the temperature dependence of the calculated Raman scattering intensity in $B_{1g}$ channel at doping level $x=0.065$ (upper panel) and $x=0.16$ (lower panel).  At the lower doping the low-frequency Raman $B_{1g}$ intensity is suppressed as the temperature decreases, whereas at the higher doping the initial slope is seen to increase as the temperature decreases again consistent with  data   \cite{Reznik93,Blumberg94,Rubhausen99,Sugai03} and with exact diagonalization of small clusters in the $t-J$ model \cite{Prelovsek96}.

\begin{figure}[htbp]
\centering
\includegraphics[angle=-0,width=0.95\columnwidth]{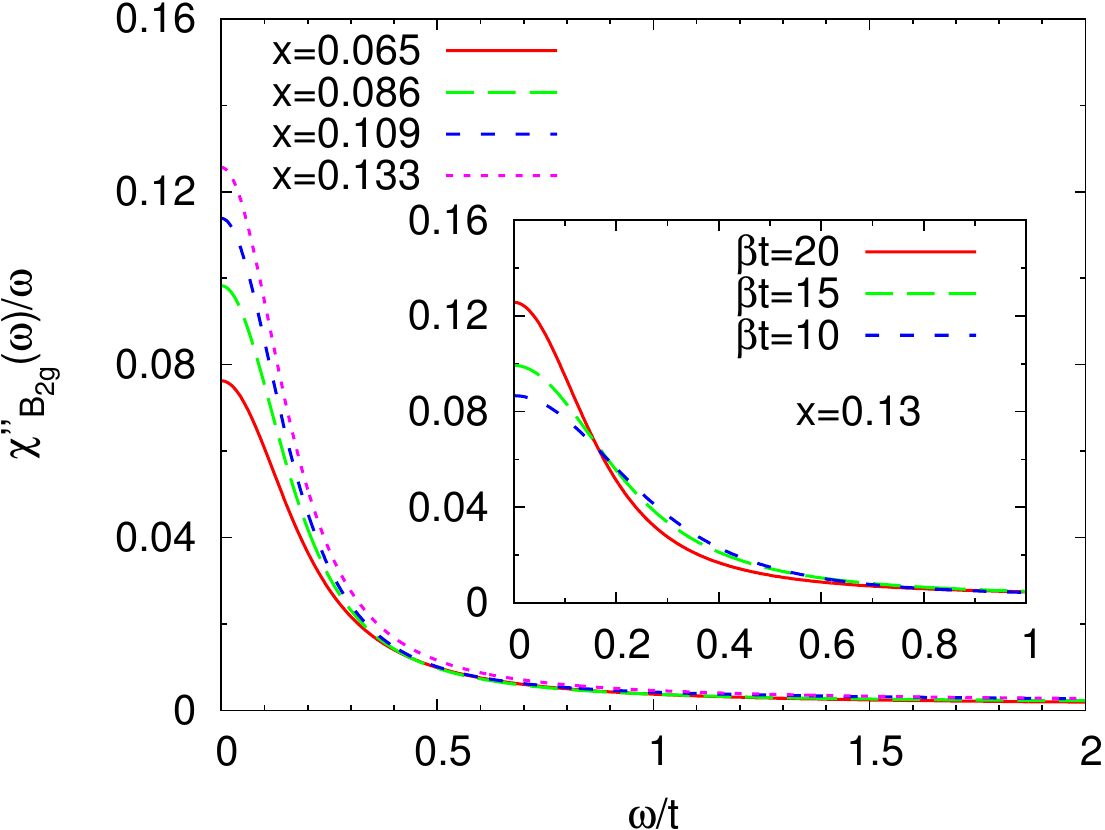}
\caption[Doping and temperature dependence of the `reinterpreted' optical conductivity in 8-site DCA]{Main panel: $B_{2g}$ Raman spectra calculated at $U=7t$ and inverse temperature $\beta=20/t\approx 200\ \text{K}$ for hole dopings indicated and presented as  ${\rm Im}\chi_{B_{2g}}/\omega$. Inset: temperature dependence of ${\rm Im}\chi_{B_{2g}}/\omega$ for hole doping $x=0.13$.}
\label{B2gByOmega}
\end{figure}

We next turn to the  $B_{2g}$ channel, which highlights the zone diagonal region of the Brillouin zone. The zone diagonal region is not affected by the pseudogap and the quasiparticle velocity is high and may therefore also be expected to dominate the optical conductivity $\sigma$. This along with a standard relation for metals motivated Ref.\cite{Shraiman90} to identify ${\rm Im}\chi_{B_{2g}}(\omega)/\omega$ with  $ {\rm Re}\sigma(\omega)$ \cite{Shraiman90}.  The main panel of Fig.~\ref{B2gByOmega}  presents the $B_{2g}$ spectra as ${\rm Im}\chi_{B_{2g}}(\omega)/\omega$. We see a `Drude' peak centered at zero frequency which grows noticeably and sharpens as doping is increased, and a broad higher-frequency continuum which is only weakly doping dependent. The inset shows that as temperature is varied the `Drude' peak  increases in height and decreases in width; there is also a small $\sim 10\%$ (not shown) increase in area with decreasing $T$.  These features bear a strong qualitative resemblance to the optical conductivity data taken in the high-$T_c$ cuprate superconductors \cite{Orenstein90,Basov05}. 

In this paper we have developed a new method  for treating the vertex corrections which are essential for the computation of wide classes of experimentally relevant spectroscopies of  interacting electron systems.
We used the formalism to show that the two dimensional Hubbard model accounts for the essential features of the doping-dependent Raman spectra observed \cite{Sugai03} in high-$T_c$ copper-oxide superconductors. The formalism introduced here is straightforwardly generalizable to most other response functions, for example to the momentum dependent spin response needed for  neutron scattering, although  in the special case of the optical conductivity   the Ward identity issues discussed in Ref.~\cite{Lin09} create complications which are not yet resolved. The  additional computational burden  of our methods scales as a power law in  cluster size, whereas the equilibrium dynamical mean-field computations themselves are limited by an exponential barrier imposed by the fermion sign problem and Hilbert space size. Therefore we expect that as computers grow more powerful, our methods can be applied to essentially any case for which an equilibrium DMFT solution is feasible, opening new directions for the theoretical understanding of correlated electron materials.

{\bf Acknowledgments} This research was supported by  NSF-DMR-1006282.  The quantum Monte Carlo calculations were performed using a code based on the ALPS \cite{ALPS20} library. We used resources of the Center for Nanophase Materials Sciences, which is sponsored at Oak Ridge National Laboratory by the Division of Scientific User Facilities, U.S. Department of Energy and of the National Energy Research Scientific Computing Center, which is supported by the Office of Science of the U.S. Department of Energy under Contract No. DE-AC02-05CH11231.

\bibliography{refs_twoparticle}

\end{document}